\ifx\pdfminorversion\undefined\else\pdfminorversion=7\fi
\ifx\pdfsuppresswarningpagegroup\undefined\else\pdfsuppresswarningpagegroup=1\fi
\documentclass[a4paper,11pt]{article}
\usepackage{pos}

\usepackage{cancel}
\usepackage{slashed}
\usepackage{ucs}
\usepackage[T1]{fontenc}
\usepackage{graphicx}
\graphicspath{{figs/}}

\def\color#1{\relax}
\colorlet{yellow}{yellow!70!black}

\newcommand\Tr{\mathop{\rm Tr}}

\usepackage[capitalize]{cleveref}

\makeatletter
\newcount\Hy@sublinkcounter
\newcommand\phantomsubsection{%
  \Hy@GlobalStepCount\Hy@sublinkcounter
  \xdef\@currentHref{subsection*.\the\Hy@linkcounter.\the\Hy@sublinkcounter}%
  \Hy@raisedlink{\hyper@anchorstart{\@currentHref}\hyper@anchorend}}
\expandafter\renewcommand\expandafter\phantomsection\expandafter
  {\phantomsection\global\Hy@sublinkcounter0\relax}

\@ifundefined{texorpdfstring}
{\newcommand\texorpdfstring[2]{#1}}{}
\@ifundefined{unichar}
{\newcommand\myunichardef[3]{\expandafter\providecommand\csname text#1\endcsname
                            {#2}}}
{\newcommand\myunichardef[3]{\expandafter\providecommand\csname text#1\endcsname
                            {\unichar{"#3}}}}
\makeatother

\myunichardef{composetilde}{\string~}{0303}
\myunichardef{composemacron}{bar}{0305}
\myunichardef{subscriptthree}{\string_3}{2083}
\myunichardef{superscripttwo}{\string^2}{00B2}

\renewcommand\subsection[1]{\relax}

\title{nEDM from the theta-term and chromoEDM operators}
\ShortTitle{\texorpdfstring{nEDM from \(\Theta\) and qcEDM}{nEDM from \texttheta\ and qcEDM}}

\author*[a]{Tanmoy Bhattacharya}
\author[b]{Vincenzo Cirigliano}
\author[a]{Rajan Gupta}
\author[a]{Emanuele Mereghetti}
\author[c]{Boram Yoon}

\affiliation[a]{Los Alamos National Laboratory,\\
  MS B285, P.O. Box 1663, Los Alamos, NM 87545-0285, USA}

\affiliation[b]{Physics Department, University of Washington,\\
  3910 15th Avenue NE, Seattle, WA 98195-1560, USA}

\affiliation[c]{NVIDIA Corporation, Santa Clara, CA 95050, USA}

\emailAdd{tanmoy@lanl.gov}
\emailAdd{cirigv@uw.edu}
\emailAdd{rg@lanl.gov}
\emailAdd{emereghetti@lanl.gov}
\emailAdd{byoon@nvidia.com}

\abstract{In a previous work, we showed that unresolved excited state
contaminations provide a major source of systematic uncertainty in the
calculation of the nucleon electric dipole moment due to the QCD
topological term theta. Here we extend  the calculation
 to the quark chromo-electric dipole moment operator (qcEDM). 
 We also show quantitatively the impact of the 
mixing of the qcEDM with lower-dimensional operators on the lattice.
Finally, we present preliminary results from a unitary clover-on-clover calculation for the QCD topological term.}

\FullConference{%
  The 39th International Symposium on Lattice Field Theory (Lattice2022),\\
  8-13 August, 2022 \\
  Bonn, Germany 
}

\begin{document}
\maketitle

\section{Introduction}
\label{sec:intro}

\subsection{BSM Operators}

In spite of its remarkable success in explaining almost all current experiments, the standard model of particle physics~\cite{Gaillard1999} cannot be complete---it cannot explain the observed cosmology.  In particular, it cannot give rise to a universe with the observed vast excess of matter over antimatter~\cite{Kolb:1990vq}.  Any initial excess would either have been diluted to nothing due to inflation, or would have to be large enough to make current inflationary models impossible~\cite{Coppi:2004za}.  As pointed out by Sakharov~\cite{Sakharov:1967dj}, one of the three conditions for a  dynamical generation of this asymmetry is violation of the symmetry under simultaneous charge-conjugation (C) and parity (P) flip, which is called the CP-symmetry. The CP-violation (CPV) in the standard model due to the phase in the quark-mixing matrix is too small to generate enough matter~\cite{Shaposhnikov:1987tw,Farrar:1993sp}.
Thus, we need CPV beyond the standard model (BSM), and if the source of this violation couples to quarks and gluons, we generically expect static electric dipole moments (EDM) of hadrons with non-zero spin.  %In this work, we shall discuss the EDMs of nucleons (nEDM) due to BSM CPV couplings to heavy particles.
We will work with a low energy effective theory with operators of 
dimension 4 and higher obtained by integrating out heavy BSM degrees of freedom
%.  By naturalness, the expected contribution of these operators decreases as a power dictated by their dimension, hence,
and here consider only operators up to dimension 6.\looseness-1

At dimension 4, we need only consider CPV due to the topological term in QCD, \(G_{\mu\nu}\tilde G^{\mu\nu}\).  By the singlet axial anomaly, this is related to the phase of the quark-mass determinant, which we write symbolically as \(m \bar\psi \gamma_5 \psi\), and vanishes when any quark mass is zero.
%Order-of-magnitude estimates show that an \(O(1)\) coefficient for the topological term gives rise to nEDM that are \(10^9\) times the observed bounds, so this term must be unnaturally small. This can be accommodated by invoking a Peccei-Quinn symmetry which dynamically relaxes the coefficient to a value set by other sources of low-energy CPV in the theory, and hence generically small. This, however, forces us to calculate the effect of the topological term when we are considering BSM CPV---any such CPV will have a direct effect on nEDM, but also an indirect effect by the induced topological term.
%
At dimension 5, we have two operators, both of which arise only after the Higgs field acquires a vacuum expectation value, \(v_{\rm EW}\), from operators that are dimension 6 when the weak interaction symmetry group is unbroken.  Their coefficients are, therefore, expected to be suppressed by \(v_{\rm EW}/M^2_{\rm BSM}\), where \(M_{\rm BSM}\) is the BSM energy scale of the heavy particles integrated out.  These two operators are the quark electric dipole moment (qEDM), \(\bar\psi\Sigma_{\mu\nu}\tilde F^{\mu\nu}\psi\), and quark chromo-electric dipole moment (qcEDM), \(\bar\psi\Sigma_{\mu\nu}\tilde G^{\mu\nu}\psi\).
At dimension 6, we encounter the gluon chromo-electric dipole moment operator (gcEDM), also called the CPV Weinberg 3-gluon operator, \(G_{\mu\nu}G_{\lambda\nu}\tilde G_{\mu\lambda}\), and CPV four-Fermi operators with various Lorentz and flavor structures.
%Of these, lattice calculations of nEDM due to qEDM are mature and will not be discussed any further.  Preliminary analysis of the effects of the Weinberg operator are available, but plagued by mixing with lower dimensional operators. The four-fermi operators have, so far, not been studied on the lattice.  In this work, we focus on the remaining two operators: the QCD topological term and the qcEDM operator.

\subsection{States}

\begin{figure}
\begin{center}
  \includegraphics[width=0.45\textwidth]{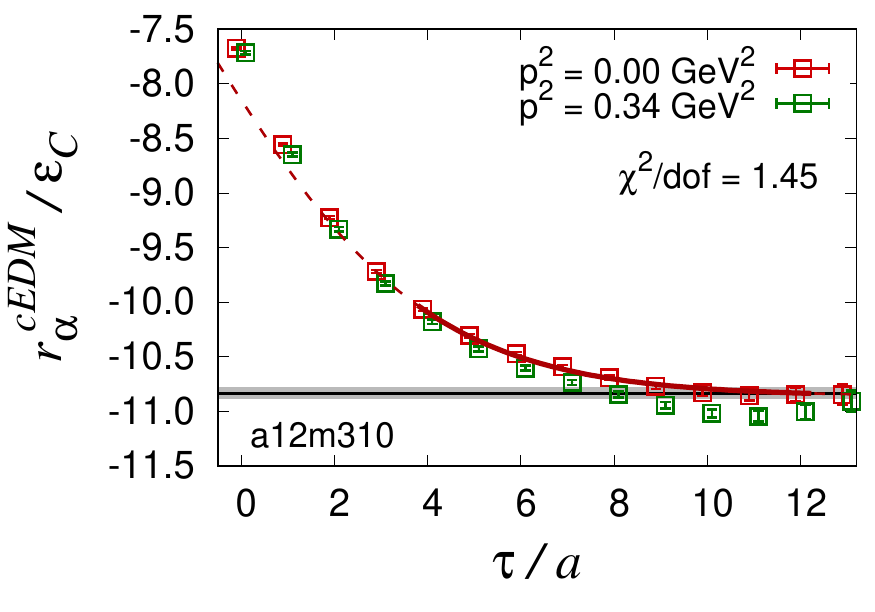}
  \includegraphics[width=0.45\textwidth]{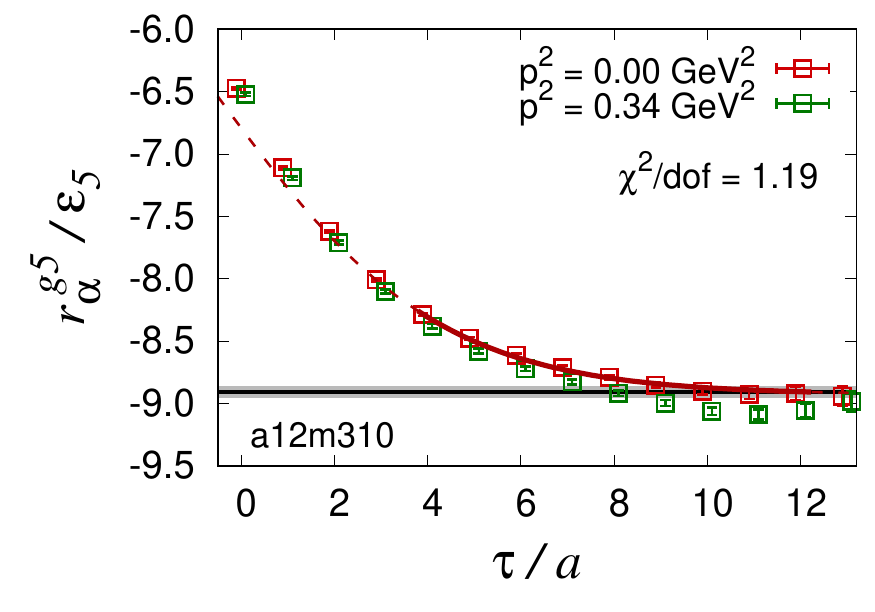}
\end{center}
\vspace*{-1.5\baselineskip}
\caption{Determination of \(\alpha_N\) arising from the isovector qcEDM and \(\bar\psi\gamma_5\psi\) operators on the \(a\approx0.12~\rm fm\), \(M_\pi\approx310~\rm MeV\) ensemble (called a12m310 henceforth), see \cref{eq:ralpha} for details.}
\label{fig:ralpha}
\end{figure}

Before calculating the contribution of these CPV operators to nEDM, we note that CP transformations of elementary particle states needs careful definition~\cite{Bhattacharya:2021lol}.  The point is that in the Lehmann-Symanzik-Zimmermann (LSZ) reduction, we first need to define asymptotic states that behave as free particles in the relevant weak limit---which implies they have all the symmetries of the noninteracting limit, including \(P\). If the interaction does not have these symmetries, the symmetry generator, however, varies with the asymptotic state, and is a property of the dynamics. %In particular, the generators for the P-transformation of a quark cannot be used in a straightforward way to construct the generator for the P-transformation of a nucleon unless parity is a good symmetry of the dynamics.
Nevertheless, any interpolating operator \(\hat N\) for the asymptotic nucleon state constructed to have the proper Lorentz properties can always be rotated to \( \hat {\tilde N} = e^{-i\alpha_N \gamma_5} \hat N \) to obtain the standard parity operator on the asymptotic state. Furthermore, the \(\alpha_N\) can be chosen real if interactions have \(PT\) symmetry, as we assume here. A nonperturbative determination of \(\alpha_N\) can be obtained from the nucleon two-point function:
\begin{equation}
 \lim_{\tau\to\infty} \left[r_\alpha(\tau) \equiv \frac{\Im\Tr\gamma_5(1+\gamma_4)\langle \hat N(\tau)\overline {\hat N}(0)\rangle}
    {\Re\Tr(1+\gamma_4)\langle \hat N(\tau)\overline {\hat N}(0)\rangle} \right]\,.
\label{eq:ralpha}
\end{equation}
In \cref{fig:ralpha}, we show an example of the determination of this phase for the isovector qcEDM operator and the isovector \(\bar\psi\gamma_5\psi\) operator with which it mixes.

\subsection{Form Factors}

When working with the rotated fields, \(\hat{\tilde N}\), the operations \(C\), \(P\), and time reversal, \(T\), are represented by the standard operators, and the
Dirac \(F_1\), Pauli \(F_2\), electric dipole \(F_3\), and anapole \(F_A\) form factor decomposition of the vector current between nucleon states takes the standard form.
%\begin{align}
% \langle N | V_\mu(q) | N \rangle & = 
%    \overline {u}_N \left[ 
%          \gamma_\mu\;F_1(q^2) + i \frac{[\gamma_\mu,\gamma_\nu]}2 q_\nu\; \frac{F_2(q^2)}{2 m_N} \right.%\nonumber\\
 %%&\qquad
 %         {} + \Orange{(2 i\,m_N \gamma_5 q_\mu - \gamma_\mu \gamma_5 q^2)\;\frac{F_A(q^2)}{m_N^2}} \nonumber\\[1\jot]
 %&\qquad\qquad\left.
 %         {} + \Green{\frac{[\gamma_\mu, \gamma_\nu]}2 q_\nu \gamma_5\;\frac{F_3(q^2)}{2 m_N}} \right] u_N\,
 %\end{align}
 %where \(| N\rangle\) is an asymptotic state created by \(\hat{\bar{\tilde N}}\) acting on the vacuum.
The Sachs electric \(G_E \equiv F_1 - (q^2/4M_N^2) F_2\) and magnetic \(G_M \equiv F_1 + F_2\) form factors are also related to these as usual. In particular, for the electromagnetic current, \(G_E(0) = F_1(0)\) is the electric charge which is \(0\) for the neutron and \(1\) for the proton; \(G_M(0)/2 M_N = F_2(0) / 2 M_N\) is the magnetic dipole moment; \(F_3(0)/2 M_N\) is the CPV electric dipole moment and \(F_A\) violates PT, and is \(0\) in our calculation.\looseness-1

%% \section{Lattice Calculation}
%% \label{sec:latt}

%% \subsection{Technique}

The nEDM is thus obtained from the electric dipole moment form factor $F_3$ at zero momentum transfer, which requires calculating the matrix element of the electric current that is the source of the electromagnetic field in the presence of CPV. When the CPV is the qEDM, the result is just the tensor charge; for the other operators, {{it is the usual vector electromagnetic quark bilinear.} }

The inclusion of CPV due to qcEDM operator is straightforward using the Schwinger source method~\cite{Bhattacharya:2016oqm}.  Since the operator is a local quark bilinear, it can be included by modifying the propagator.
%% \begin{equation}
%%    \left.\Dslash + m - \frac r2 D^2 + c_{sw} \Sigma^{\mu\nu} G_{\mu\nu}\right. \mathbin{\Red{\longrightarrow}}
%%    \left.\Dslash + m - \frac r2 D^2 + \Sigma^{\mu\nu} ( c_{sw} G_{\mu\nu} + i \epsilon \Green{\tilde G_{\mu\nu}} )\right.\,,\!\!\!\!
%% \end{equation}
In the isovector case, the Fermion determinant is not modified.
%% \begin{align}
%% &\frac {\det ( \Dslash + m - \frac r2 D^2 + \Sigma^{\mu\nu} ( c_{sw} G_{\mu\nu} + i \epsilon \Green{\tilde G_{\mu\nu}} )}
%%           {\det ( \Dslash + m - \frac r2 D^2 + c_{sw} \Sigma^{\mu\nu} G_{\mu\nu} )} \\
%% =&
%% \exp \Tr \ln \left[1 + i \epsilon\, \Green{\Sigma^{\mu\nu} \tilde G_{\mu\nu}} ( \Dslash + m - \frac r2 D^2 + c_{sw} \Sigma^{\mu\nu} G_{\mu\nu} )^{-1}\right]\,.
%% %\\
%% % &\approx&
%% % \exp \left[ i \epsilon \Tr \Green{\Sigma^{\mu\nu} \tilde G_{\mu\nu}} ( \Dslash + m - \frac r2 D^2 + c_{sw} \Sigma^{\mu\nu} G_{\mu\nu} )^{-1}\right]\,.
%% \end{align}
The $\Theta$-term and the gluon chromo-EDM operators are purely disconnected contributions and their calculation ends up being a factor reweighting the vector-current 3-point function~\cite{Bhattacharya:2022whc}.

\section{Quark Chromoelectric Dipole Moment}
\label{sec:qcedm}

\subsection{Three-point function}

For the qcEDM operator, the propagators are evaluated with the qcEDM operator with a small coefficient \(\epsilon\) included in the Dirac operator.
%% \begin{equation}
%% \vcenter{\hbox{\includegraphics[width=0.05\textwidth]{qEDM}}} \times 
%% \left(\;\vcenter{\hbox{\includegraphics[width=0.3\textwidth]{cEDM_3pt_disc}}} + 
%%  \vcenter{\hbox{\includegraphics[width=0.3\textwidth]{cEDM_3pt}}}\;\right)\,.
%% \end{equation}
Since this operator is dimension 5, multiple insertions bring in contributions that diverge as higher powers of the lattice spacing as we approach the continuum limit.  This necessitates us to do the calculation with \(\epsilon\) small enough to avoid the effects of these multiple insertions.  In practice, we ensure this by staying in the linear regime of \(\epsilon\)~\cite{Bhattacharya:2022whc}.

\subsection{Excited state fits}

\begin{figure}
\begin{center}
\includegraphics[page=1,viewport=40 290 390 528,clip,width=0.45\hsize]{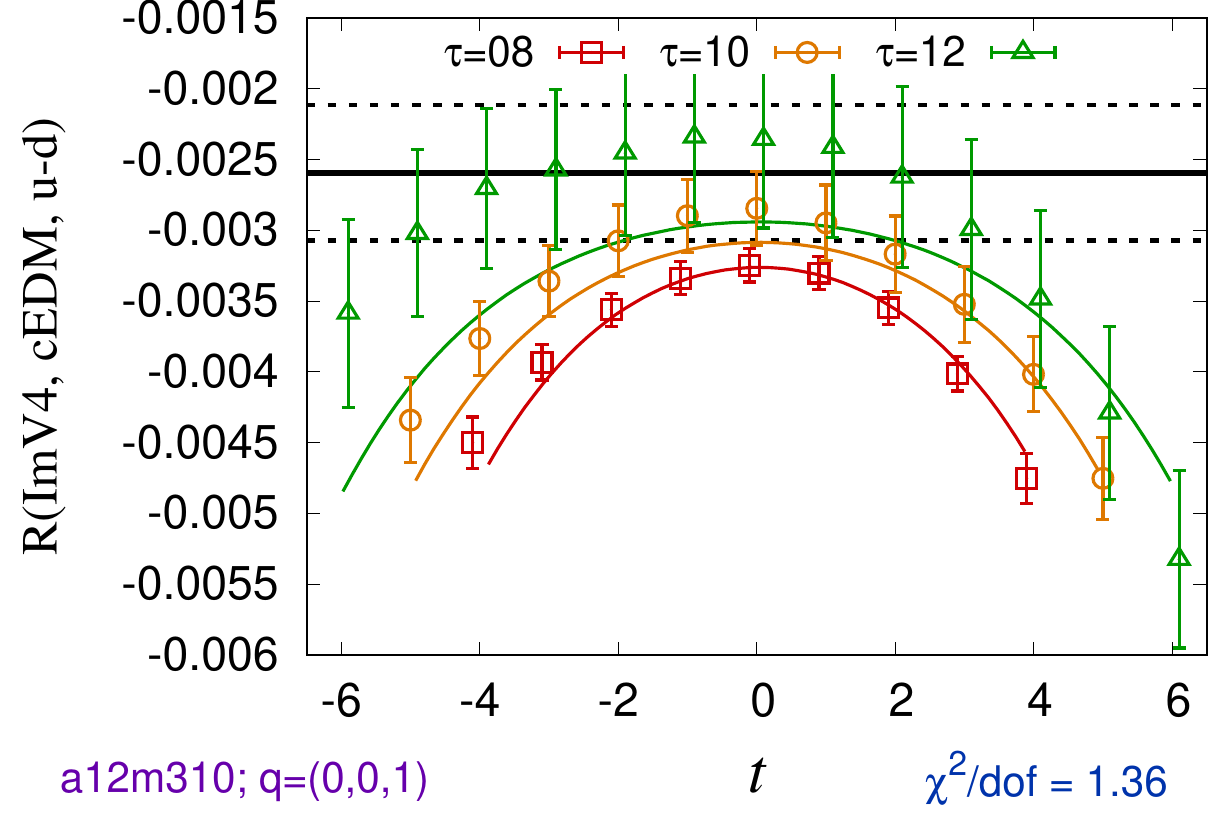}
\includegraphics[page=1,viewport=40 290 390 528,clip,width=0.45\hsize]{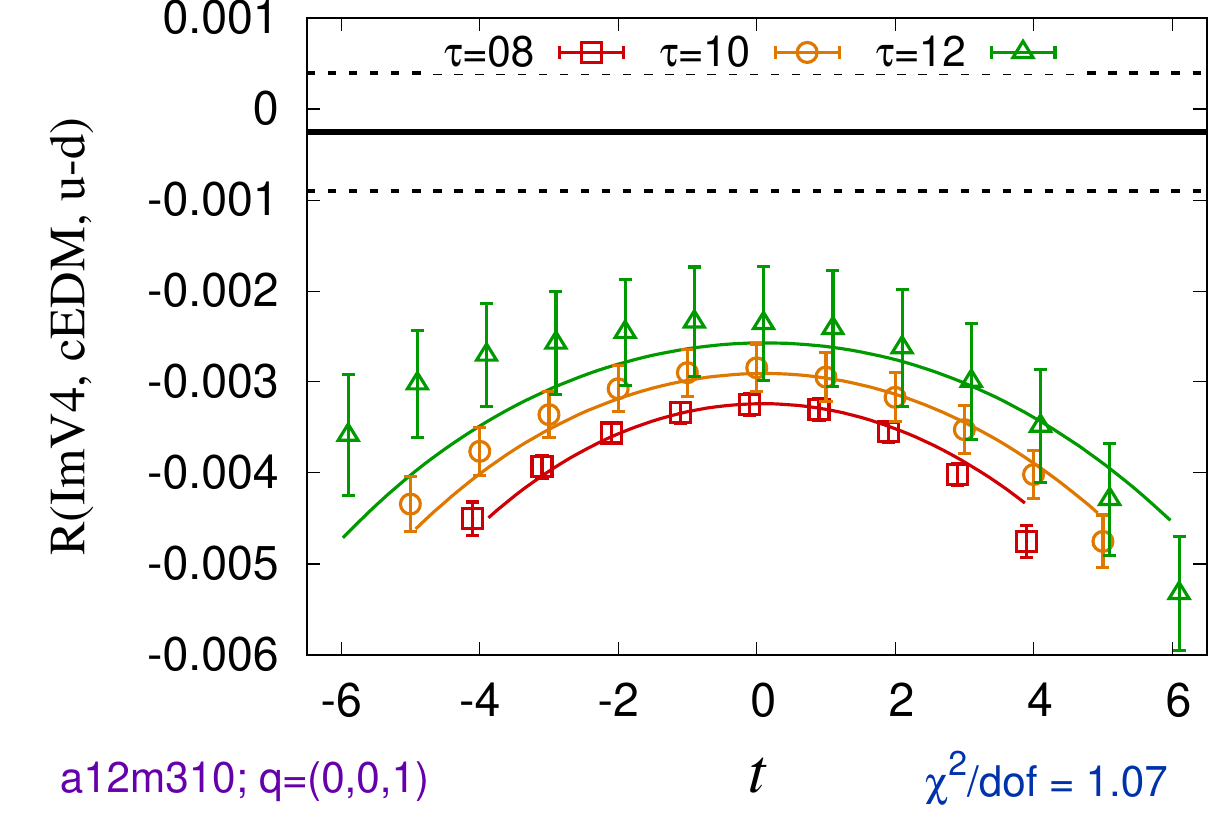}
\end{center}
\vspace*{-\baselineskip}
\caption{Example of ESC subtraction using two strategies for the 3-pt function in the presence of an isovector qcEDM CPV with nucleon source-sink separation \(\tau\) and insertion of \(V_4(q)\) at \(t\): on the left, we assume that the excited state effects are saturated by those extracted from fitting to the 2-point function; on the right, we assume that the excited state effects arise from the lowest  \(N\pi\) multihadron intermediate.}
\label{fig:ESC}
\end{figure}

The nucleon interpolating operators couple not only to the nucleon, but all single- and multiparticle states of appropriate symmetry. Traditionally, one used the 2-point functions, which have a larger signal-to-noise ratio, to obtain the spectrum, and used this in the fits to the 3-point function to extract the ground-state matrix elements.  The quality of the fits to the 3-pt functions alone are relatively insensitive to the spectrum, but the extracted matrix elements are sensitive to it.
Recently, it was discovered~\cite{Jang:2019vkm} that in some 3-point functions, the transition matrix elements between the ground and a low-lying excited state make a sizable contribution even when the fits to the 2-point function do not discover the excited state. As a result, it is worthwhile to study the sensitivity to assumption that the states that provide a leading correction in chiral perturbation theory do make a contribution to the 3-point function. In our example, the lightest such state is an \(N\pi\) multihadron state. 
In \cref{fig:ESC}, we show an example of a 3-point function that is fit almost equally well by both strategies, but where the result is very different under them.  This leads to a systematic error that is currently irreducible.

\subsection{Power divergence}

\begin{figure}
\begin{center}
  \includegraphics[width=0.4\hsize]{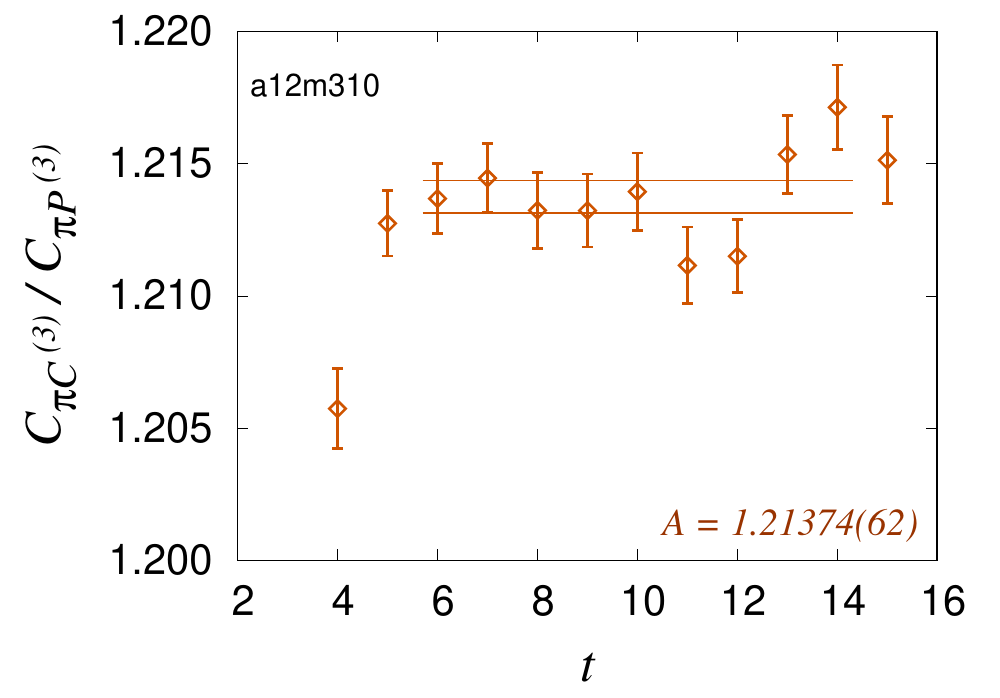}
\end{center}
\vspace*{-\baselineskip}
\caption{Example of the determination of the power-law subtraction coefficient \(A\) defined in \cref{eq:Adef} for the a12m310 ensemble.}
\label{fig:Adet}
\end{figure}

\begin{figure}
\begin{center}
\includegraphics[width=0.45\hsize]{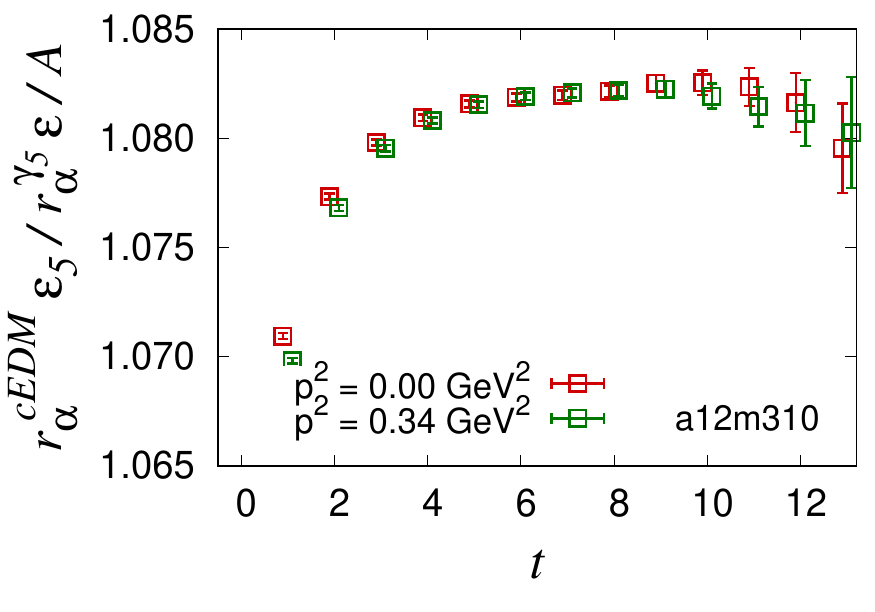}
\includegraphics[width=0.45\hsize]{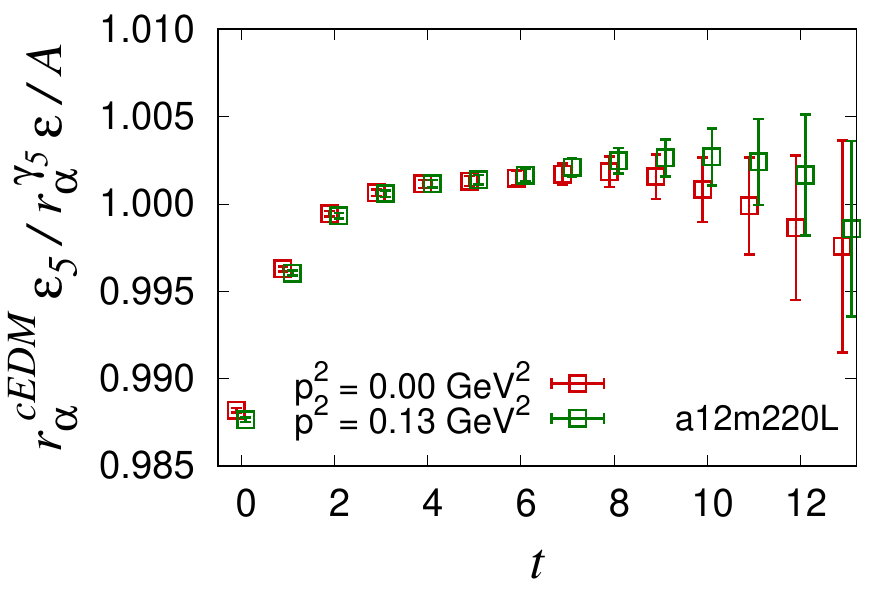}
\end{center}
\vspace*{-\baselineskip}
\caption{Ratio of \(\alpha_N\) determination from the qcEDM and \(\bar\psi\gamma_5\psi\) operators for two ensembles whose parameters are given in \cref{tab:Adet}. See \cref{eq:ratis1} for details.}
\label{fig:alpharat}
\end{figure}

Since the lattice is a hard-cutoff regularization, even the on-shell matrix elements of the qcEDM operator have power-law divergences.  On general symmetry grounds, one can define a subtracted operator \(\tilde C\) whose on-shell matrix elements diverge at most logarithmically:
\begin{equation}
  \tilde C = i \bar\psi \Sigma^{\mu\nu}\gamma_5G_{\mu\nu}T^a\psi - i\frac A{a^2}\bar\psi\gamma_5T^a\psi\,.
  \label{eq:Adef}
\end{equation}
A convenient condition for fixing \(A\) is demanding \(\left\langle\Omega\right|\tilde C\left|\pi\right\rangle=0\). In \cref{fig:Adet}, we show an example of determination of this coefficient, and in \cref{tab:Adet}, we provide the lattice parameters and the value of \(A\) for the ensembles used in our study.
This choice is especially convenient in leading order $\chi$PT, since it implies that
\begin{equation}
\alpha_N(\tilde C) \approx 0 \implies \frac1A\frac{\alpha_N(C)}{\alpha_N(\bar\psi\gamma_5\psi)} \approx 1\,.
\label{eq:ratis1}
\end{equation}
In \cref{fig:alpharat}, we show by example that this relation is true to about 10\% in our calculations.

\begin{table}
\begin{center}
     \begin{tabular}{c|cc|cc}
      Ensemble&\(c_{SW}\)&\(a\) (fm)  &\(t\)-range&\(A\)           \\ \hline
      a12m310 & 1.05094  & 0.1207(11) & 6--14     & \(1.21374(62)\)\\
      a12m220L& 1.05091  & 0.1189(09) & 7--14     & \(1.21800(33)\)\\
      a09m310 & 1.04243  & 0.0888(08) & 8--22     & \(0.99621(30)\)\\
      a06m310 & 1.03493  & 0.0582(04) &14--30     & \(0.77917(24)\)\\
    \end{tabular}
\end{center}
\vspace*{-\baselineskip}
    \caption{Lattice parameters and the value of the power-law subtraction coefficient \(A\) defined in \cref{eq:Adef}.}
    \label{tab:Adet}
\end{table}

\subsection{Multiplicative renormalization}
\begin{figure}
\begin{center}
  \includegraphics[width=0.5\hsize]{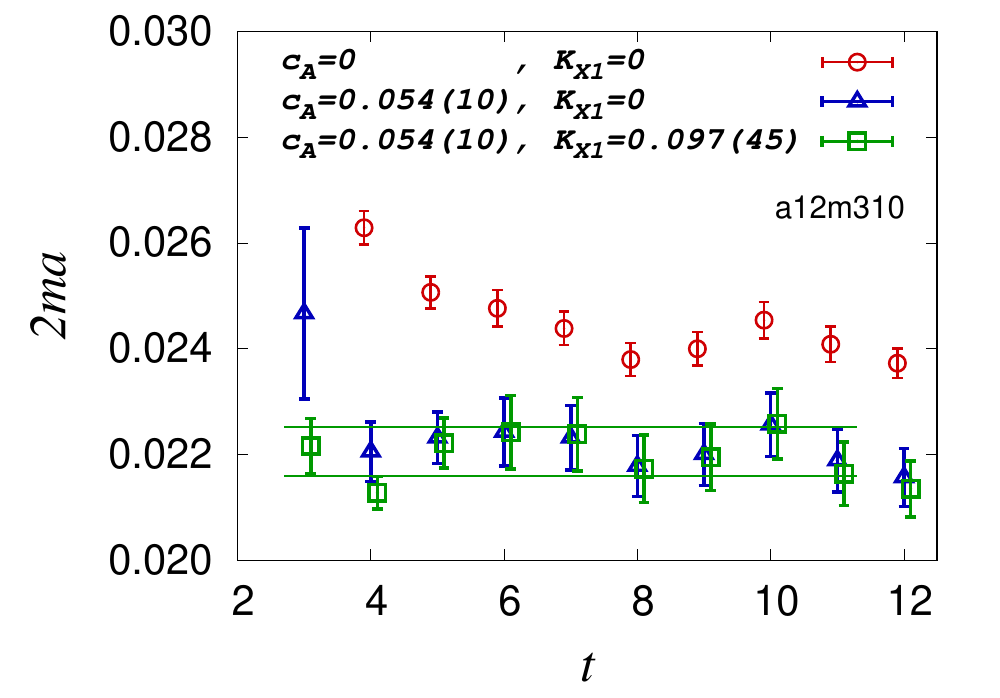}
\end{center}
\vspace*{-\baselineskip}
\caption{Determination of the nonperturbative coefficients defined in \cref{eq:AWI} for the a12m310 ensemble.}
\label{fig:xy}
\end{figure}

For our current study of isovector qcEDM operator, this power-law mixing does not lead to a divergence in the physical effects.  This is because the non-anomalous axial rotations allows the isovector pseudoscalar operator to be rotated away. The only subtlety is that with Wilson-clover Fermions used in our study, the discretization breaks the axial symmetry explicitly, and leaves behind \(O(a)\) effects.  These can be studied explicitly by writing the nonanomalous axial Ward identity
\begin{equation}
  \frac{\langle\pi \left[a\partial_\mu A^\mu - \bar c_A a^2\partial^2 P + \bar K(a^2C-AP)\right]\rangle}
  {\langle\pi P\rangle} = 2\bar m a (1 + O(a^2))\,,
  \label{eq:AWI}
\end{equation}
where \(A^\mu\) and \(P\) are the isovector axial current and pseudoscalar operators, respectively, and \(c_A\) and \(\bar K\) are nonperturbative coefficients. This equality can be used to determine \(c_A\) from the long time behavior of appropriate 2-point functions, and then \(\bar K\) from intermediate times. In \cref{fig:xy}, we show an example of the determination of these constants, and in \cref{tab:xy}, we report the values for the various ensembles.  It is important to note that there is an important interplay between two small constants: \(\bar K\), which is zero if \(c_{\rm sw}\) is nonperturbatively tuned and \(ma\), the light quark mass in the theory. \looseness-1
Furthermore, because of \cref{eq:AWI}, on-shell at zero-momentum, we have
\begin{equation}
\hbox{M.E. of } P = \hbox{M.E. of } \frac {x \equiv a^2\bar K}{y \equiv 2\bar m a + A\bar K} C\,,
\end{equation}
so that the power-law subtraction leads to an effect proportional to the qcEDM operator itself, with the proportionality constant of order unity. 

\begin{table}
\begin{center}
     {\tabcolsep=1pt
      \begin{tabular}{c|cc|cc|rrrrr}
        &\multicolumn{2}{c|}{{fit-range}}&\multicolumn{2}{c|}{{\(\chi^2/\rm d.o.f\)}}&&&&\\
      Ensemble&{\(c_A\)}&{\({\bar K}\)}&{\(c_A\)}&{\({\bar K}\)}&\multicolumn1c{\(c_A\)}&\multicolumn1c{\({\bar{K}}\)}&\multicolumn1c{\({2\bar m a}\)}&\multicolumn1c{{\(\displaystyle\frac{2 \bar m a}{K}\)}}&\multicolumn1c{{\(\displaystyle\frac{2 \bar m a} {2 m a + AK\strut}\)}}\\\hline
      a12m310  & 4--11 & 3--11 & 0.66 & 0.88 & $0.054(10)$  & $0.097(45) $ & $0.02205(46) $ & $0.23(10) $ &  $0.158(58) $ \\
      a12m220L & 4--11 & 3--11 & 2.08 & 3.09 & $0.0342(77)$ & $0.183(35) $ & $0.01152(21) $ & $0.063(12)$ &  $0.0491(86)$ \\
      a09m310  & 5--15 & 4--15 & 0.99 & 1.09 & $0.0277(40)$ & $0.047(15) $ & $0.01684(15) $ & $0.35(11) $ &  $0.263(61) $ \\
      a06m310  & 6--20 & 5--20 & 0.29 & 1.53 & $0.0093(17)$ & $0.0272(60)$ & $0.010460(37)$ & $0.385(87)$ &  $0.331(50) $ \\
    \end{tabular}}
\end{center}
\vspace*{-\baselineskip}
    \caption{The nonperturbative coefficients defined in \cref{eq:AWI} for various ensembles.}
    \label{tab:xy}
\end{table}

\subsection{Results}

\begin{figure}
\begin{center}
  \includegraphics[viewport=23 315 380 570, clip, width=0.45\hsize]{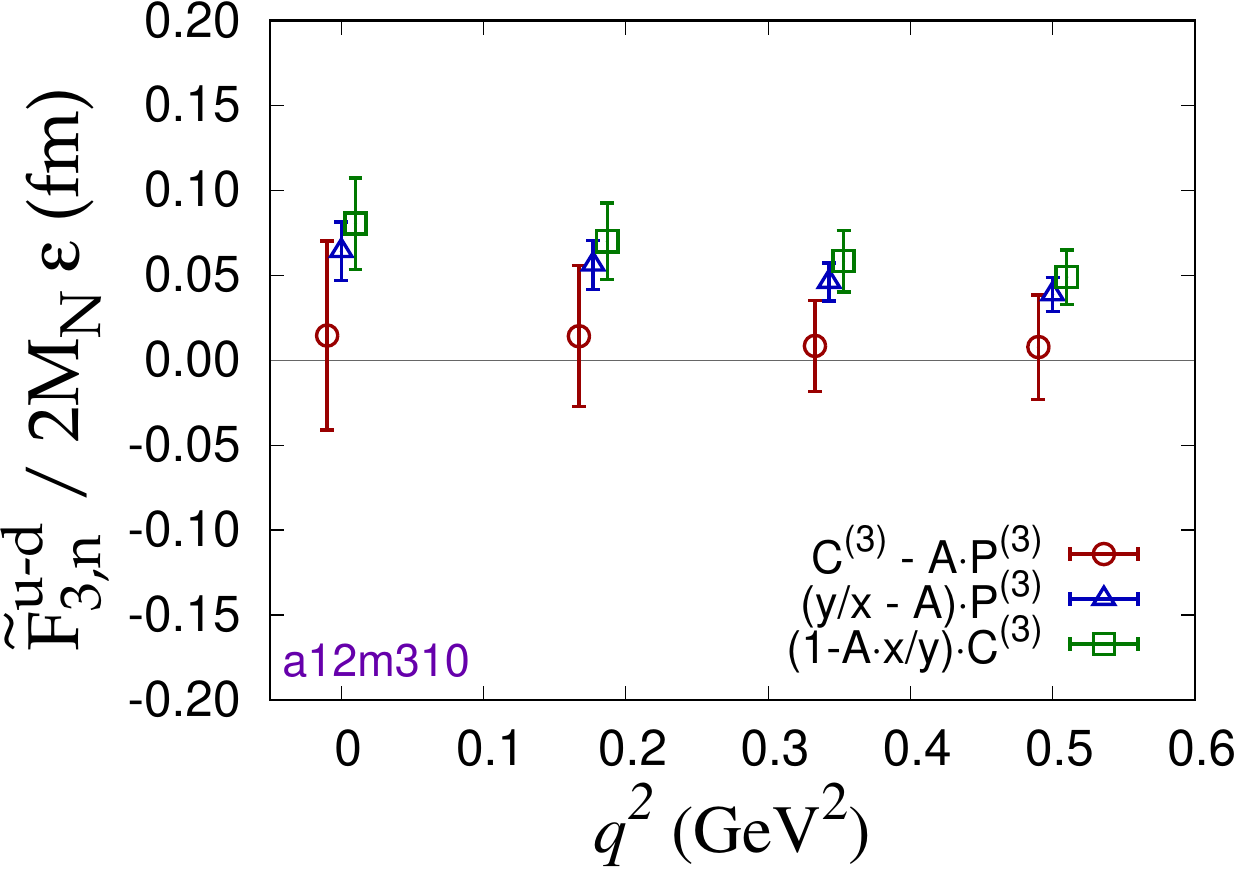}\quad
  \includegraphics[viewport=23 40 380 300, clip, width=0.45\hsize]{Ctilde_neutron_neutronESC}
\end{center}
\vspace*{-\baselineskip}
\caption{The power-law subtracted nEDM for the neutron for two ensembles.}
\label{fig:F3qcEDM}
\end{figure}

\begin{figure}
\begin{center}
  \includegraphics[width=0.47\textwidth]{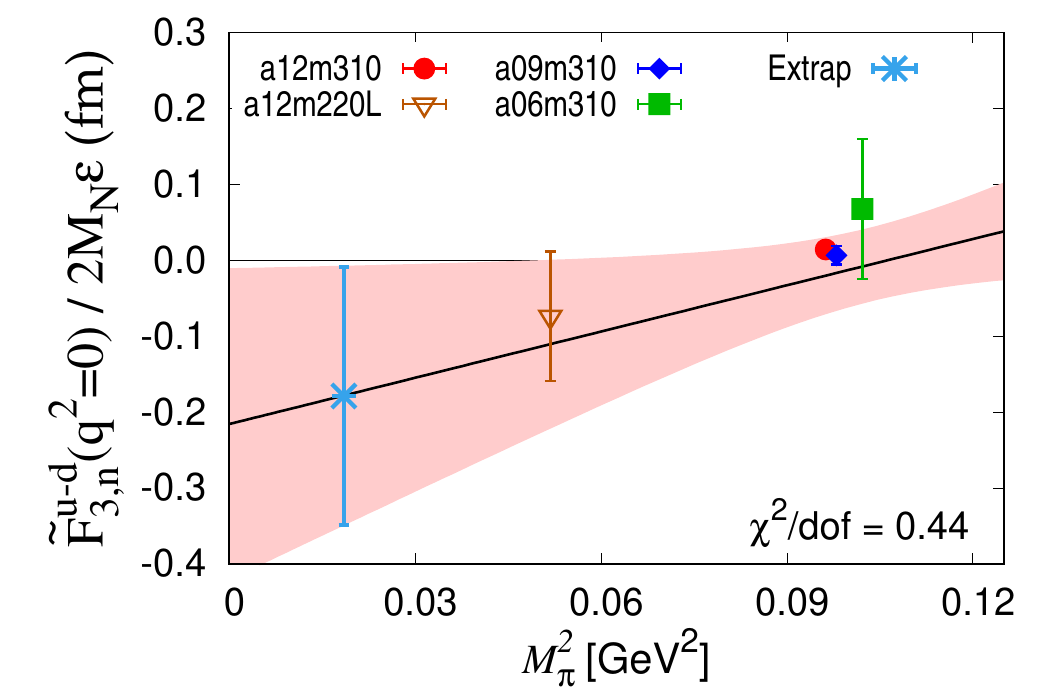}\qquad
  \includegraphics[width=0.47\textwidth]{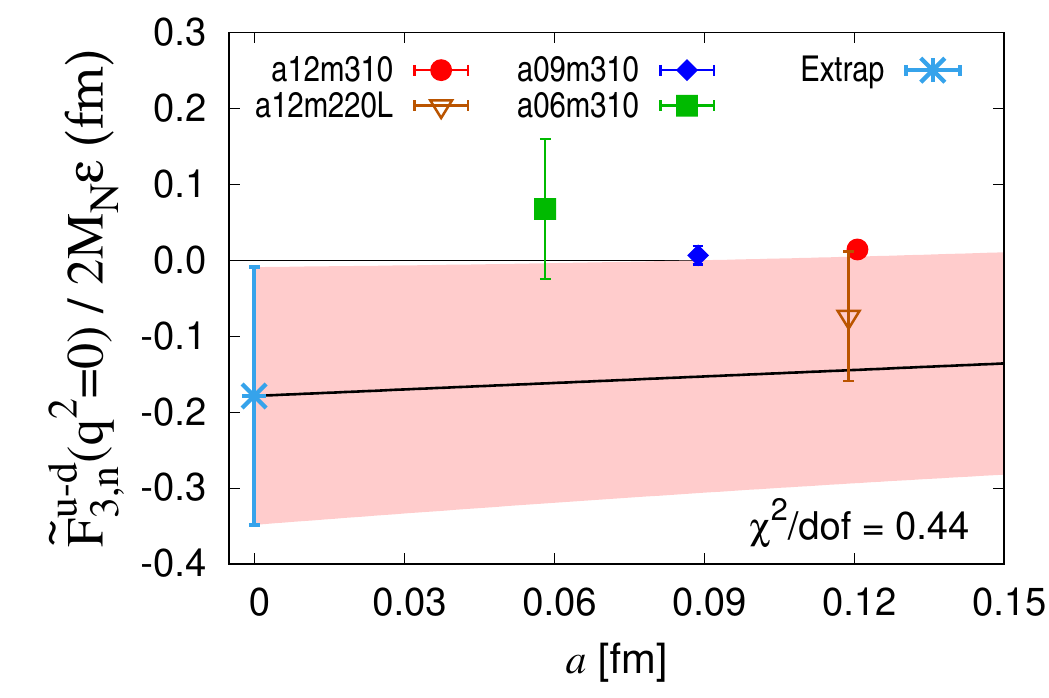}
\end{center}
\vspace*{-\baselineskip}
\caption{Chiral-continuum extrapolation for nEDM due to qcEDM.}
\label{fig:CCFV}
\end{figure}

Putting everything together, we can calculate the CPV form-factor due to the power subtracted qcEDM operator \(\tilde C\) in three ways: either by multiplicatively renormalizing the lattice \(C\) or the lattice \(P\) operator, or by explicitly subtracting the two. In \cref{fig:F3qcEDM}, we show the quality of the determination by the three methods. We note that, presumably due to the smallness of \(\bar K\) and \(am\), the \(O(a^2)\) effects neglected in \cref{eq:AWI} make a relatively large contribution and the difference between the three methods gives a large systematic uncertainty.
Ignoring these systematic uncertainties, as well as the logarithmic renormalization and mixing, the continuum-chiral extrapolation of the nEDM is shown in \cref{fig:CCFV}. We note that there is a trend towards more negative values at lower quark masses, whereas the continuum extrapolation is almost flat.

\section{QCD topological term}

\subsection{Comparing clover and HISQ sea}

\begin{table}
\begin{center}
\begin{tabular}{||l|l|l||l|l|l||}
Name&\(a\) (fm)&\(M_\pi\) (MeV)&%
Name&\(a\) (fm)&\(M_\pi\) (MeV)\\
\hline
a127m285&0.127(2)&285(3)&%
a094m270L&0.094(1)&269(3)\\
a094m220&0.094(1)&214(3)&%
a094m220s&0.0925(10)&217(3)\\
a091m170&0.091(1)&170(2)&%
a073m270&0.0728(8)&272(3)\\
a071m170&0.0707(8)&167(2)&%
a056m280&0.056(1)&281(5)
\end{tabular}
\end{center}
\vspace*{-\baselineskip}
\caption{2+1 clover-on-clover ensembles available for study. The a094m220s ensemble is at the SU(3) point, the rest have \(m_s\) close to the physical value.}
\label{tab:cloverens}
\end{table}

\begin{figure}
    \centering
    \includegraphics[width=0.45\hsize]{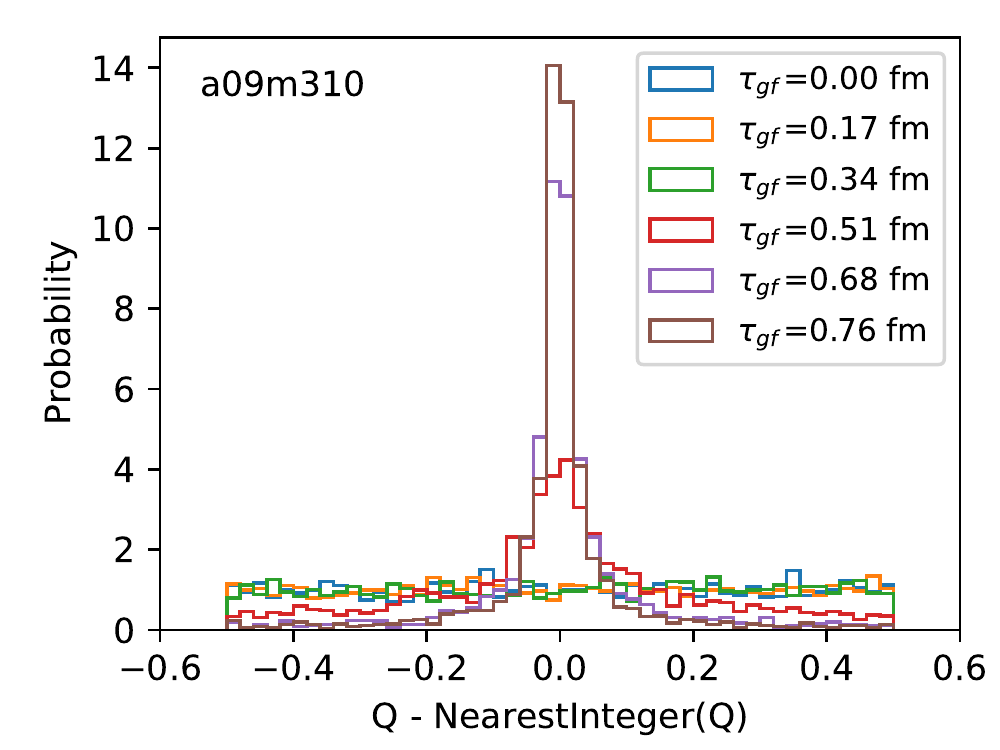}
    \includegraphics[width=0.45\hsize]{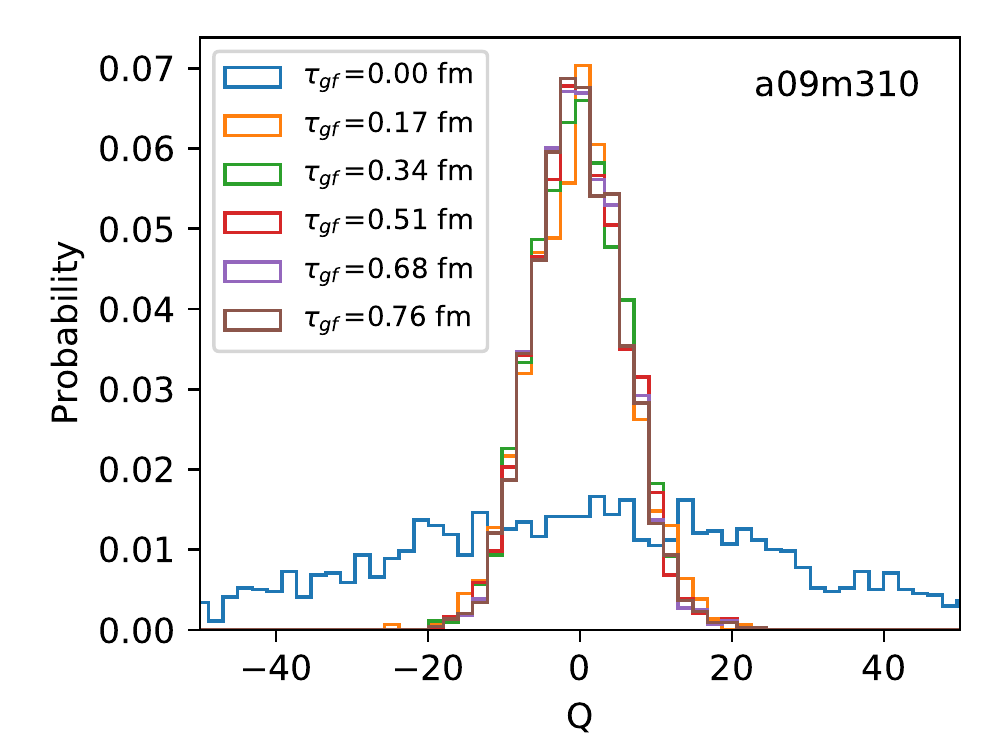}
    \includegraphics[width=0.45\hsize]{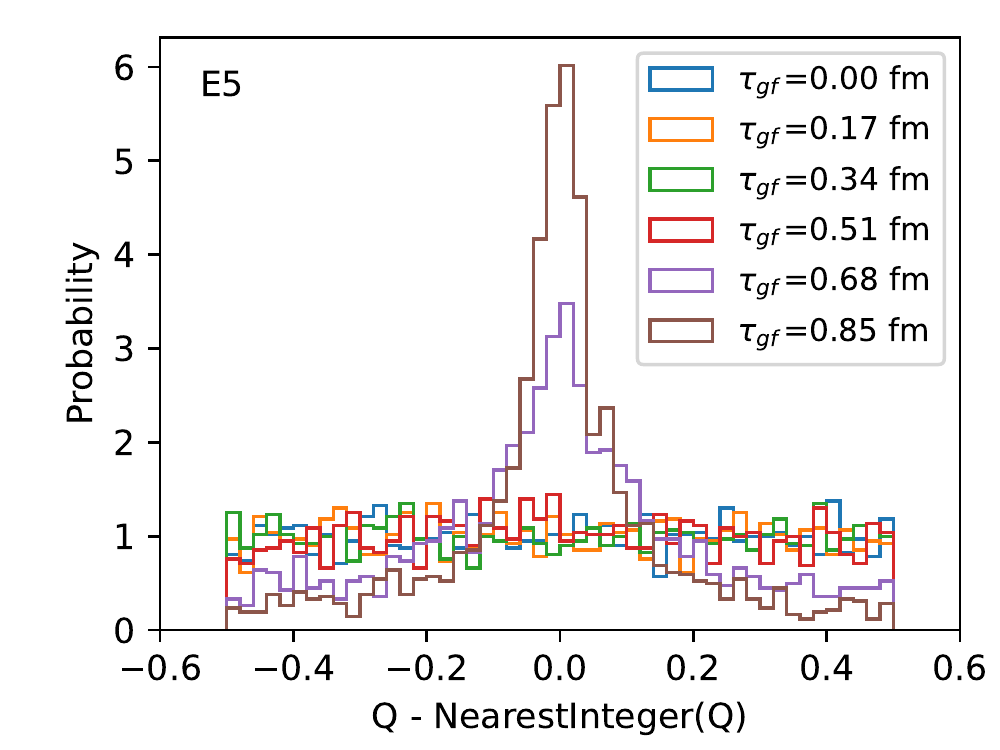}
    \includegraphics[width=0.45\hsize]{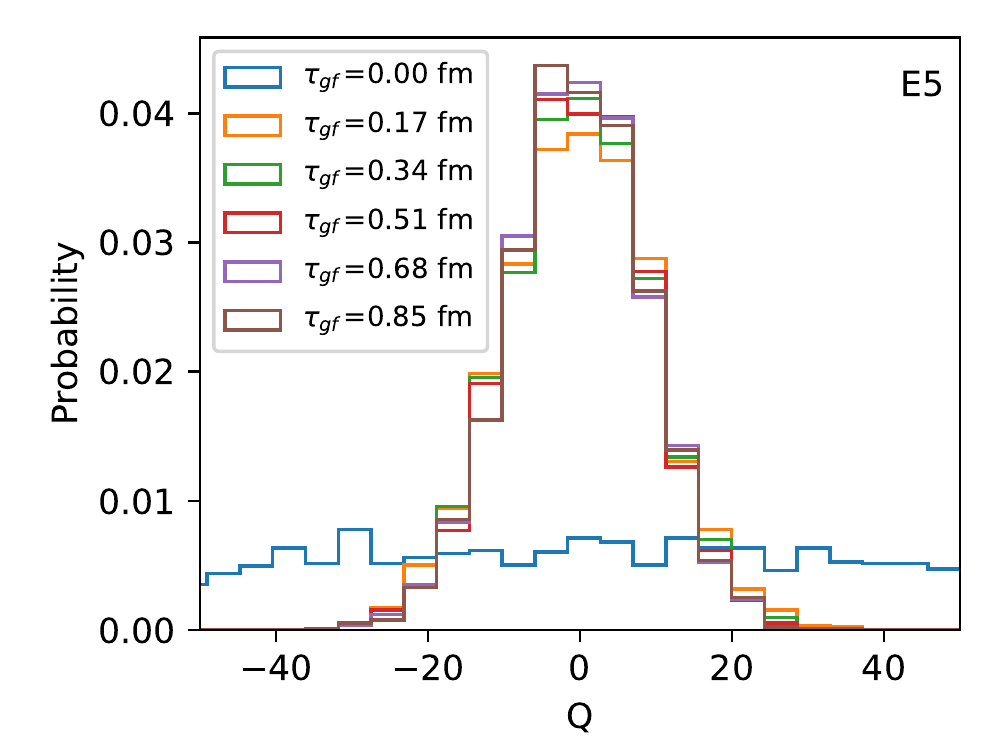}
    \caption{Topological charge (right) and its fractional part (left) as a function of gradient flow in HISQ a09m310 (top) and clover a073m270 (bottom) ensembles.}
    \label{fig:gflow}
  \end{figure}

We previously reported~\cite{Bhattacharya:2021lol} calculations of the nEDM due to the QCD topological term using a mixed action calculation with clover valence quarks on a HISQ sea, using the same ensembles presented above.  These had lattice spacings in the range 0.057--0.151~fm, pion masses in the range 128--320~MeV, and used between 550 and 2200 configurations per ensemble.  We now compare the results with a new unitary clover-on-clover calculation (see \cref{tab:cloverens}) with lattice spacings in the range 0.056--0.127~fm, pion masses in the range 167--285~MeV and using between 810 and 2100 configurations per ensemble. In \cref{fig:gflow}, we show that the behavior of the charge under gradient flow is similar in both the calculations.

\subsection{Topological Susceptibility}

\begin{figure}
    \centering
    \includegraphics[width=0.45\hsize]{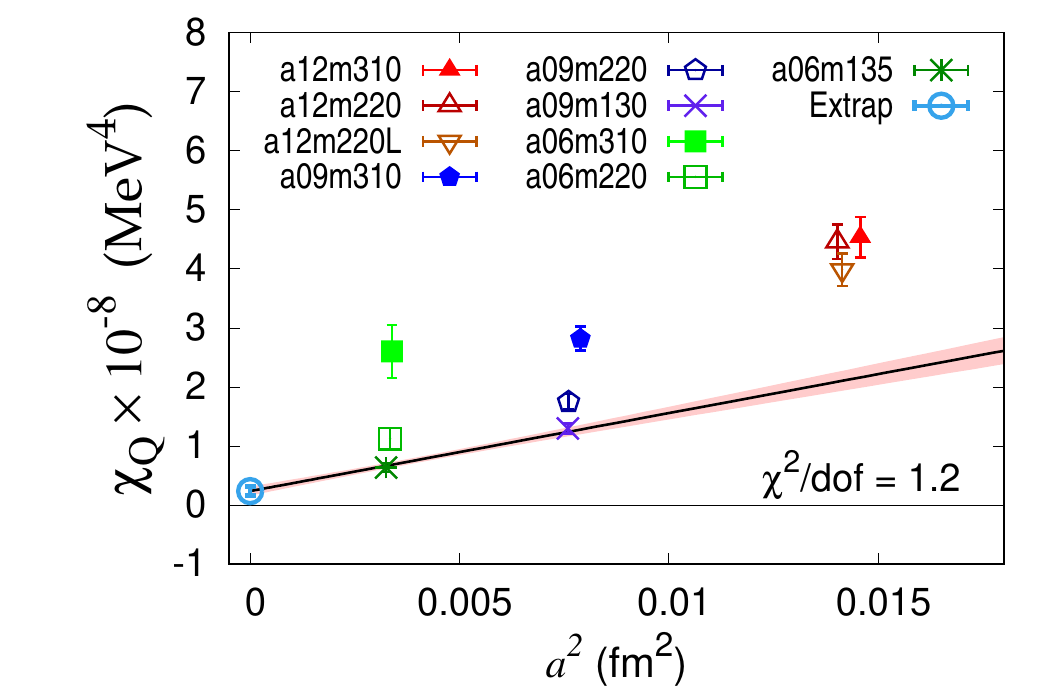}
    \includegraphics[width=0.45\hsize]{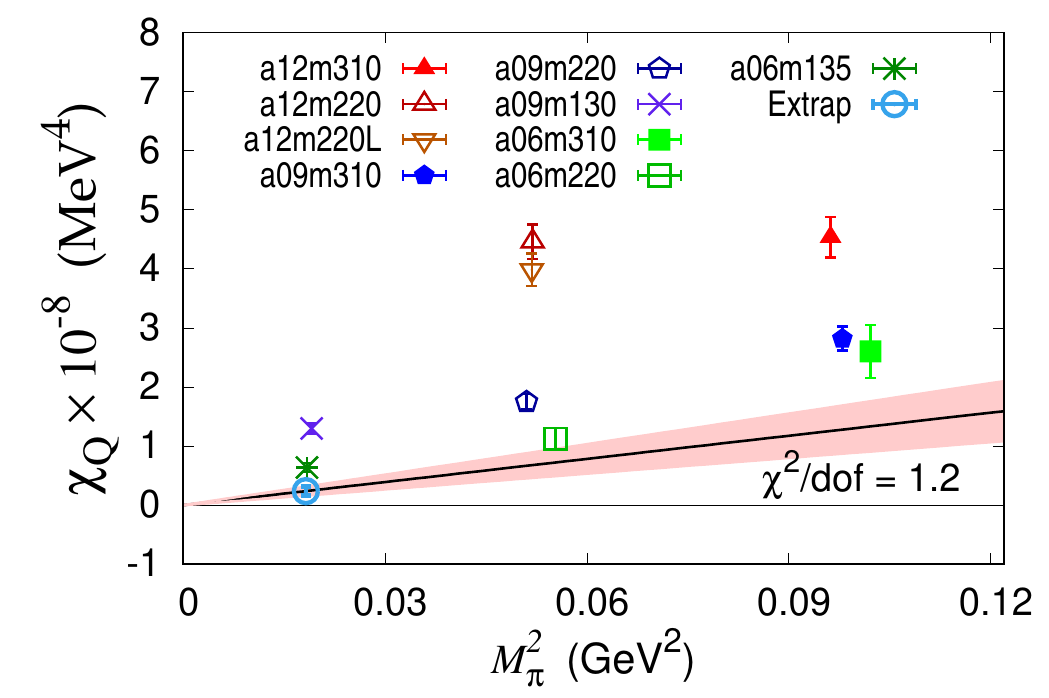}
    \includegraphics[width=0.45\hsize]{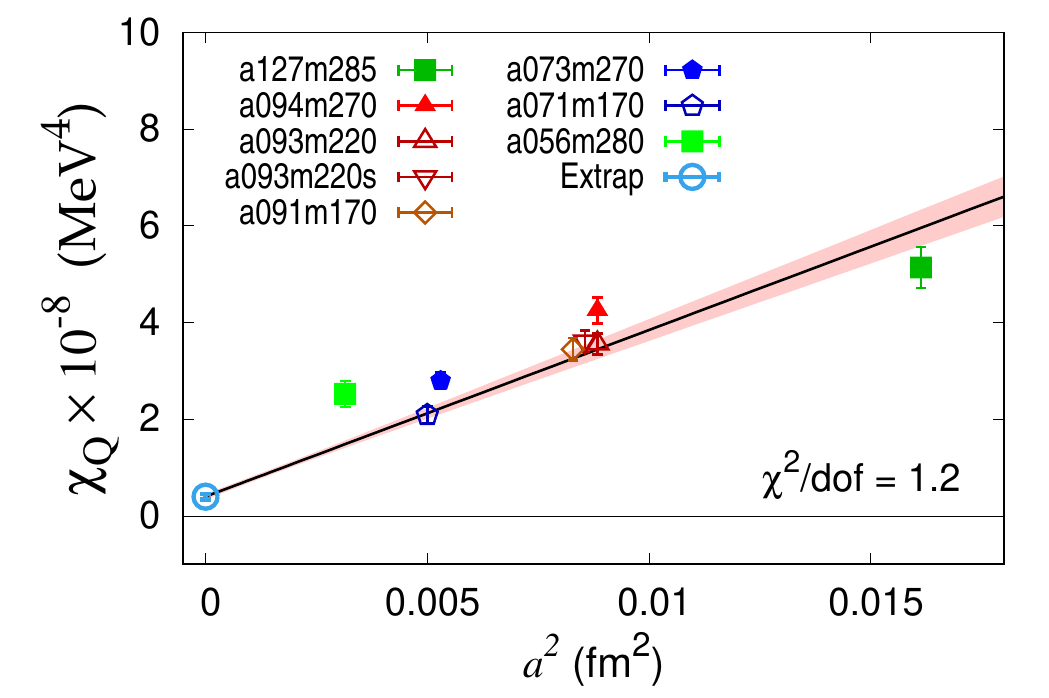}
    \includegraphics[width=0.45\hsize]{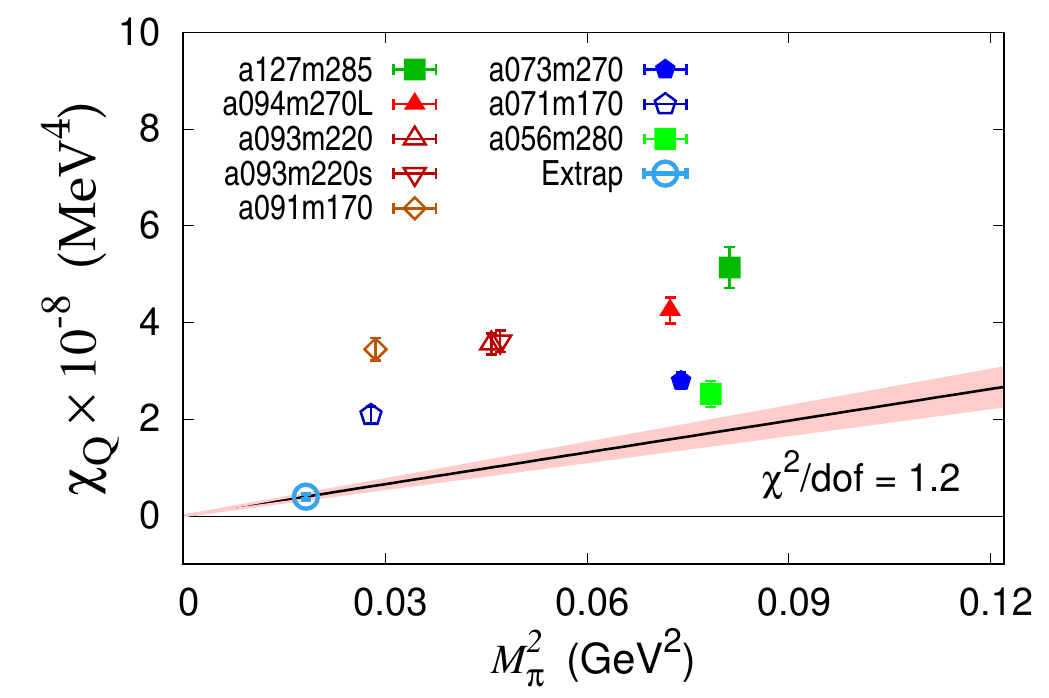}
    \caption{The chiral-continuum extrapolation of \(\chi_Q\) in the HISQ (top) and clover (bottom) ensembles.}
    \label{fig:susceptibility}
  \end{figure}

 As shown in \cref{fig:susceptibility}, the new preliminary result for the topological susceptibility, \([79.5(3.0) \hbox{ MeV}]^4\), from the clover lattices is very similar to \(\chi_Q = [66(9)(4) \hbox{ MeV}]^4\) obtained from the HISQ lattices.  Both are in good agreement with the expectation from $\chi$PT:
      \begin{equation}
        \frac{1}{\chi_Q} = \frac{1}{\chi_Q^{\textrm{quench.}}} + \frac{4}{M_\pi^2 F_\pi^2} \left(1-\frac{M_\pi^2}{3M_\eta^2}\right)^{-1} \implies{}
               \chi_Q    = \big[79 \textrm{ MeV}\big]^4
      \end{equation}

\subsection{ESC and \texorpdfstring{\(Q^2\)}{Q\textsuperscripttwo} extrapolation}

\begin{figure}
    \centering
    \includegraphics[width=0.45\textwidth]{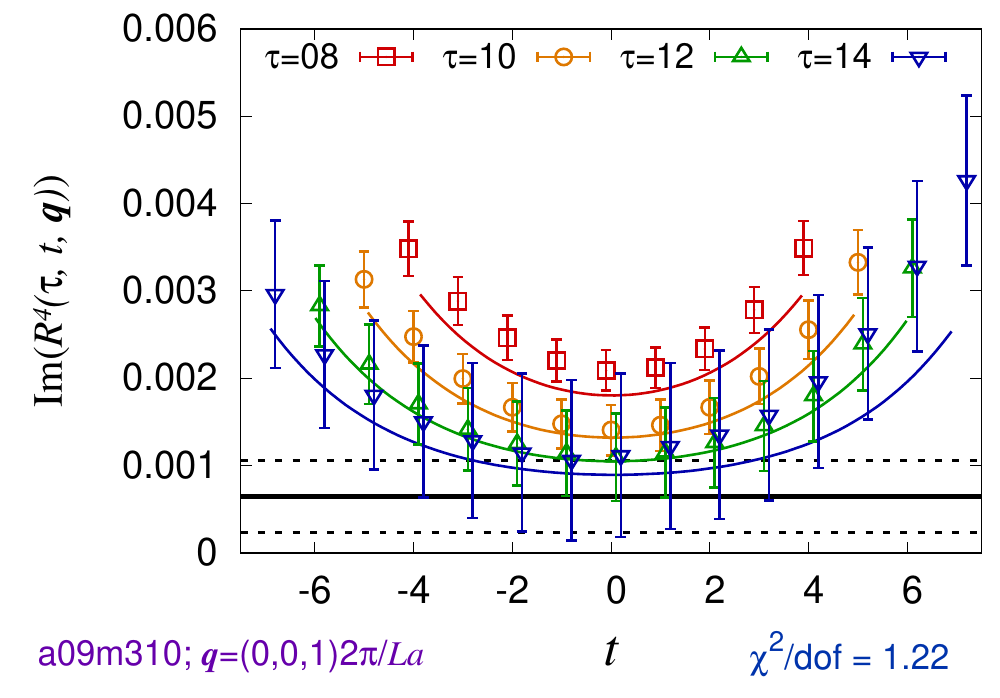}
    \includegraphics[width=0.45\textwidth]{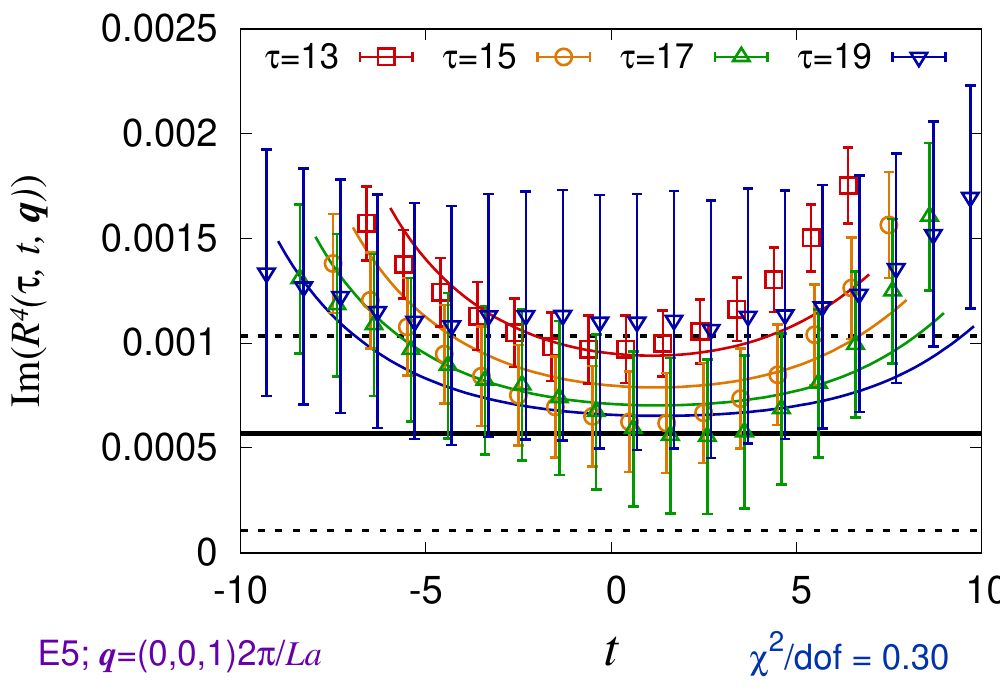}
    \caption{Example excited-state fits for the HISQ a09m310 (left) and clover a073m270 (right) ensembles.}
    \label{fig:ESCtheta}
  \end{figure}

\begin{figure}
    \centering
    \includegraphics[width=0.49\textwidth]{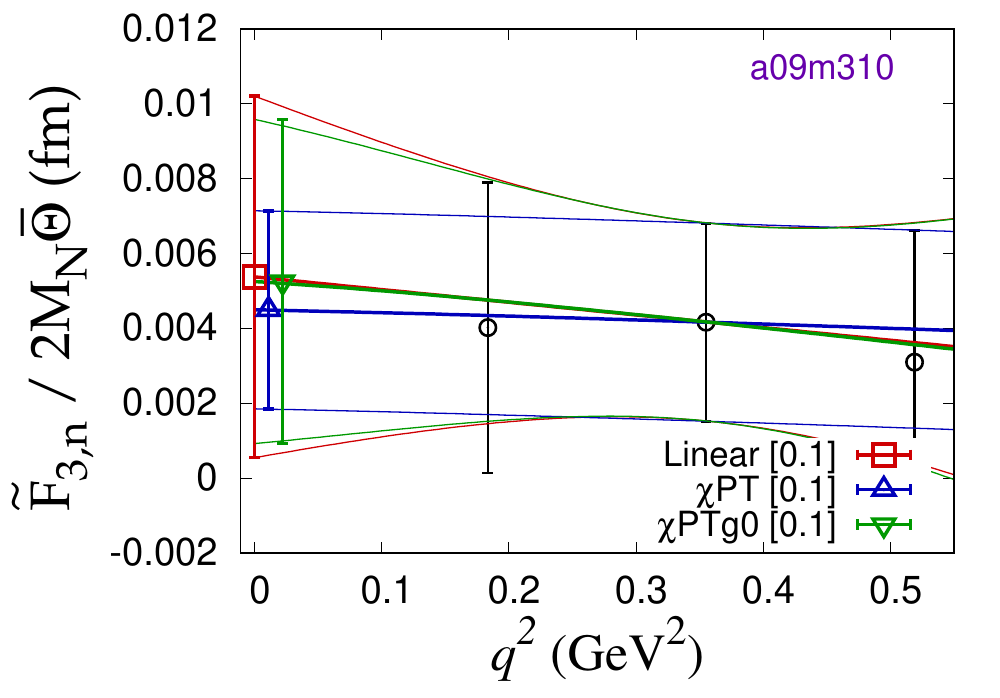}
    \includegraphics[width=0.49\textwidth]{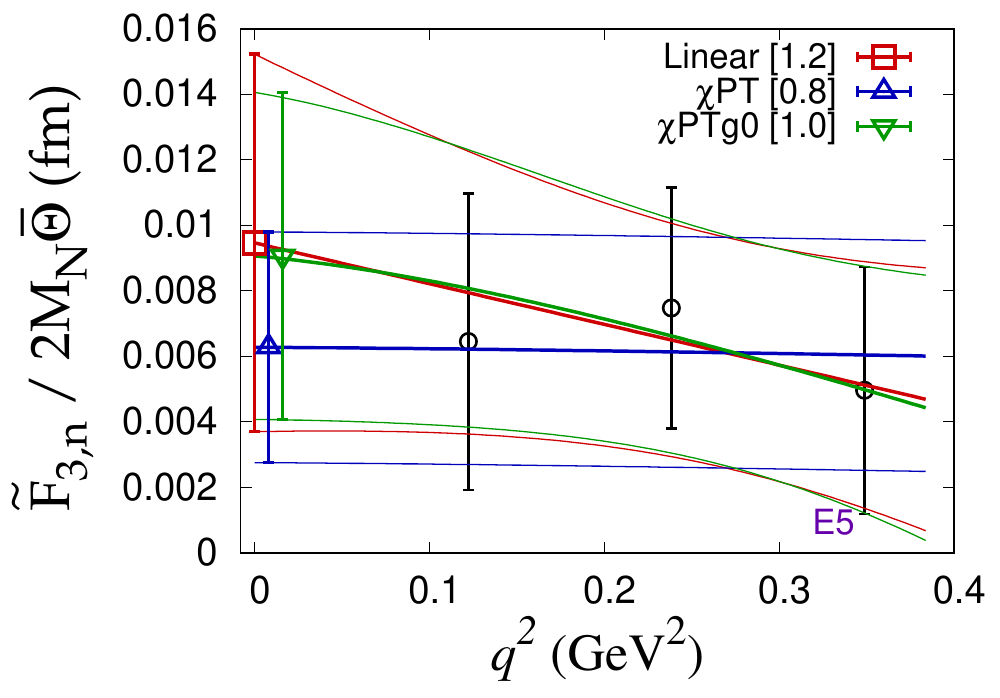}
    \caption{\(Q^2\) extrapolations for the HISQ a09m310 ensemble (left) and clover a073m270 ensemble (right).}
    \label{fig:Q2theta}
  \end{figure}

In \cref{fig:ESCtheta}, we compare the excited-state fits in the two formulations, and \cref{fig:Q2theta} shows the \(Q^2\) extrapolations. There is a qualitative agreement between the two formulations, and the errors are still large in both.
\subsection{Simultaneous fit}%
\begin{figure}
  \centering
    \includegraphics[width=0.45\textwidth]{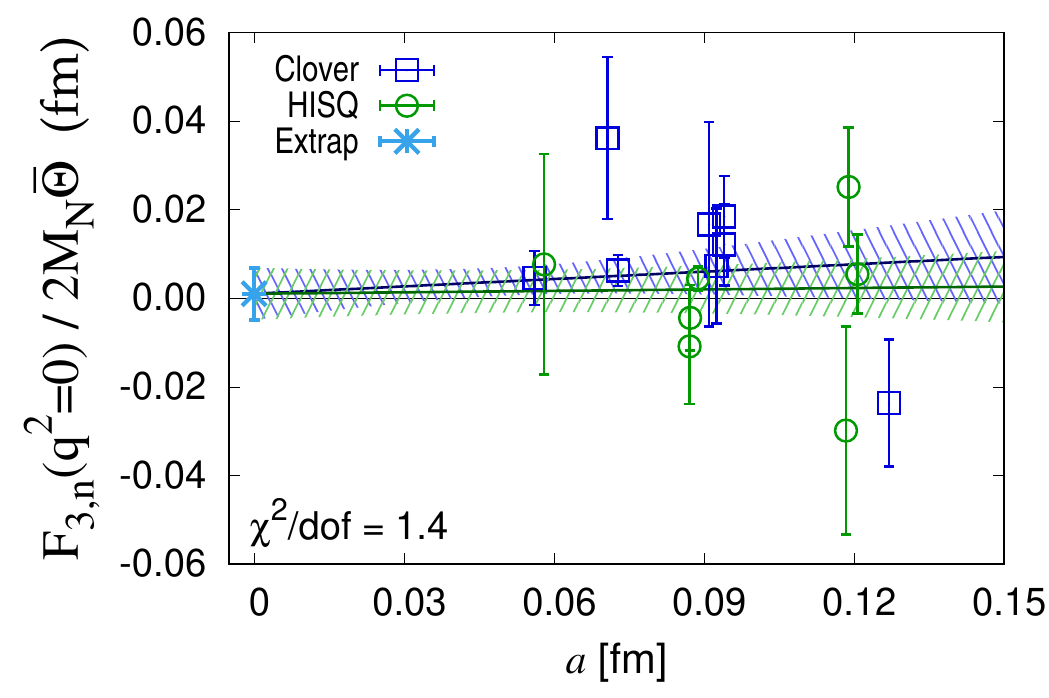}
    \quad
    \includegraphics[width=0.45\textwidth]{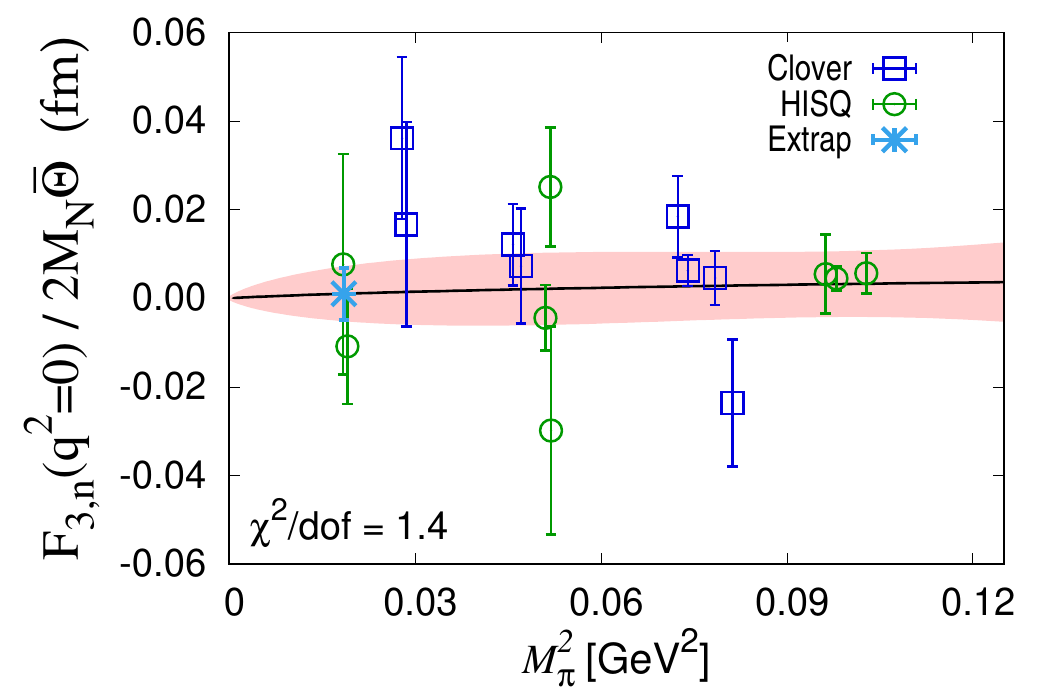}
    \caption{Simultaneous chiral-continuum fit of the clover and HISQ results.}
    \label{fig:CCFVtheta}
  \end{figure}

Finally, in \cref{fig:CCFVtheta}, we show our preliminary results for nEDM per unit qcEDM after the simultaneous extrapolation of both the clover-on-clover and clover-on-HISQ data to the continuum $a\rightarrow 0$ and physical pion mass $M_\pi \rightarrow 135\textrm{MeV}$:
      \begin{align}
        d_N = c_1 M_\pi^2 + c_2 M_\pi^2 \log\left(\frac{M_\pi^2}{M_N^2}\right) + {c_3^{\textrm{HISQ}}a} + {c_3^{\textrm{Clover}}a}
            \longrightarrow 0.0010(59)\,
      \end{align}
 where only the statistical error has been included.

\section{Future}
\label{sec:fut}

\subsection{Next steps in progress}

Several improvements are currently being worked on.  As we showed, the subtraction of the power divergence leads to uncertainties arising from the smallness of both the light quark mass and the chiral symmetry breaking. A calculation with the qcEDM operator smoothed in the gradient flow scheme~\cite{Luscher:2010iy}, will allow \(a\to0\) limit at fixed physical smearing \(t\). Since the axial Ward identity is unbroken in this limit~\cite{Luscher:2013cpa}, 
%the isovector pseudoscalar operator mixing can be completely rotated away, and the uncertainty from this source 
this uncertainty can be removed.
In addition, there is a logarithmic mixing between the nEDM obtained from the qcEDM and qEDM operators that can be perturbatively evaluated.  Since the effect of the qEDM operator is already known to very high accuracy~\cite{Park:2021ypf}, one can subtract this effect without affecting the final precision.\looseness-1

We thank the MILC collaboration~\cite{Bazavov:2012xda},  for providing the HISQ lattices. The calculations used the CHROMA software suite~\cite{Edwards:2004sx}. Simulations were carried out at (i) the NERSC supported by DOE under Contract No. DE-AC02-05CH11231;  (ii) the USQCD collaboration resources funded by DOE HEP, and (iii) Institutional Computing at Los Alamos National Laboratory. This work was supported by LANL LDRD program and TB and RG were also supported by the DOE HEP  under Contract No. DE-AC52-06NA25396.

\bibliographystyle{JHEP}
\let\oldbibitem\bibitem
\def\bibitem#1\emph#2,{\oldbibitem#1}
\let\oldthebibliography\thebibliography
\renewcommand\thebibliography[1]{\oldthebibliography{#1}%
                                 \itemsep0pt\parskip0pt\relax}
\bibliography{main}
\end{document}